\documentclass[prb]{revtex4}

\usepackage{graphicx}
\usepackage{amssymb}

\begin{document}

\title{Network Behavior in Thin Film Growth Dynamics}

\author{T. Karabacak}
\affiliation{Department of Applied Science, University of Arkansas at Little Rock, Little Rock, AR 72204, USA}

\author{H. Guclu}
\affiliation{School of Mathematical Sciences, Rochester Institute of Technology, Rochester, NY 14623, USA}

\author{M. Yuksel}
\affiliation{Department of Computer Science and Engineering, University of Nevada Ð Reno, Reno, NV 89557, USA}


\begin{abstract}

We present a new network modeling approach for various thin film growth techniques that incorporates re-emitted particles due to the non-unity sticking coefficients. We model re-emission of a particle from one surface site to another one as a network link, and generate a network model corresponding to the thin film growth. Monte Carlo simulations are used to grow films and dynamically track the trajectories of re-emitted particles. We performed simulations for normal incidence, oblique angle, and chemical vapor deposition (CVD) techniques. Each deposition method leads to a different dynamic evolution of surface morphology due to different sticking coefficients involved and different strength of shadowing effect originating from the obliquely incident particles. Traditional dynamic scaling analysis on surface morphology cannot point to any universal behavior. On the other hand, our detailed network analysis reveals that there exist universal behaviors in degree distributions, weighted average degree versus degree, and distance distributions independent of the sticking coefficient used and sometimes even independent of the growth technique. We also observe that network traffic during high sticking coefficient CVD and oblique angle deposition occurs mainly among edges of the columnar structures formed, while it is more uniform and short-range among hills and valleys of small sticking coefficient CVD and normal angle depositions that produce smoother surfaces.


\end{abstract}

\date{\today}

\maketitle


\begin{section}{Introduction}

Thin film coatings have been the essential components of various devices in industries including microelectronics, optoelectronics, detectors, sensors, micro-electro-mechanical systems (MEMS), and more recently nano-electro-mechanical systems (NEMS). Commonly employed thin film deposition \cite{1,2,3} techniques are \textit{thermal evaporation}, \textit{sputter deposition}, \textit{chemical vapor deposition} (CVD), and \textit{oblique angle deposition}. Different than others, oblique angle deposition technique \cite{4,5,6,7,8,9,10,11} is typically used for the growth of nanostructured arrays of rods and springs through a physical self-assembly process. In many applications, it is often desired to have atomically flat thin film surfaces. However, in almost all of the deposition techniques mentioned above, the surface morphology generates a growth front roughness. The formation of growth front is a complex phenomenon and very often occurs far from equilibrium. When atoms are deposited on a surface, atoms do not arrive at the surface at the same time uniformly across the surface. This random fluctuation, or noise, which is inherent in the process, may create the surface roughness. The noise competes with surface smoothing processes, such as surface diffusion (hopping), to form a rough morphology if the experiment is performed at either a sufficiently low temperature or at a high growth rate.

Due to its intractability, a conventional statistical mechanics treatment cannot be used to describe the complex phenomenon of surface morphology formation in thin film growth. About two decades ago, a dynamic scaling approach \cite{12,13} was proposed to describe the morphological evolution of a growth front. Since then, numerous modeling and experimental works have been reported based on this dynamic scaling analysis \cite{2,3,14}. On the other hand, there has been a significant discrepancy among the predictions of these growth models and the experimental results published \cite{15,16,17}. For example, various growth models have predictions on the dynamic evolution of the root-mean-square roughness (RMS), \cite{18} which is defined as $w(t)=\sqrt{[h(r,t)-\left< h\right>]^{2}}$, where $h(r,t)$ is the height of the surface at a position $r$ and time $t$, and $\left<h\right>$ is the average height at the surface. In most of the growth phenomena, the RMS grows as a function of time in a power law form, \cite{2,3,14,19} $w\sim t^{\beta}$ , where $\beta$ is the ``growth exponent'' ranging between 0 and 1. $\beta = 0$ for a smooth growth front and $\beta = 1$ for a very rough growth front (the RMS could be as large as the film thickness). Figure \ref{fig1} shows a collection of experimental $\beta$ values reported in the literature \cite{20} and compares to the predictions of growth models. Theoretical predictions of growth models in dynamic scaling theory basically fall into two categories. One involves various surface smoothing effects, such as surface diffusion, which lead to $\beta \leqslant 0.25$. \cite{2,3,14,19} The other category involves the shadowing effect (which originates from the preferential deposition of obliquely incident atoms on higher surface points and always occurs in sputtering, CVD, and oblique angle deposition) during growth which would lead to $\beta = 1$. \cite{21} However, it can be clearly seen in Fig. \ref{fig1} that experimentally reported values of growth exponent $\beta$ are far from agreement with the predictions of these growth models. \cite{20} Especially, sputtering and CVD techniques are observed to produce morphologies ranging from very small to very large $\beta$ values indicating a ``non-universal'' behavior.

\begin{figure}[htb]
\includegraphics[scale=0.5]{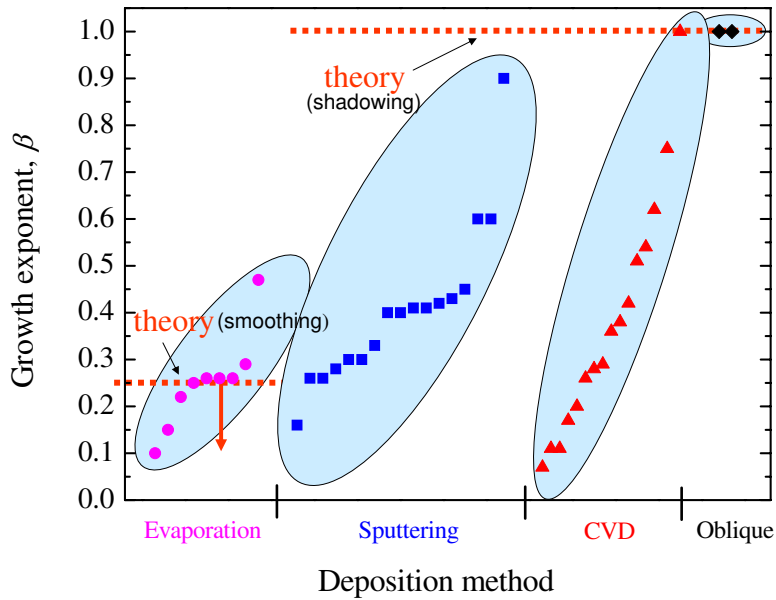}
\caption{(Color online) A survey of experimentally obtained values of growth exponent $\beta$ reported in the literature \cite{20} for different deposition techniques is compared to the predictions of common thin film growth models in dynamic scaling theory. Root-mean-square roughness (RMS) grows as a function of time in a power law form, $w\sim t^{\beta}$ , where $\beta$ is the ``growth exponent'' ranging between 0 and 1. $\beta = 0$ for a smooth growth front and $\beta = 1$ for a very rough growth front.}
\label{fig1}
\end{figure}

Only recently, it has been recognized that in order to better explain the dynamics of surface growth one should take into account the effects of both ``shadowing'' and ``re-emission'' processes. \cite{14,16,17,22,23} As illustrated in Fig. \ref{fig2}, during deposition, particles can approach the surface at oblique angles and be captured by higher surface points (hills) due to the shadowing effect. This leads to the formation of rougher surfaces with columnar structures that can also be engineered to form ``nanostructures'' under extreme shadowing conditions, as in the case of oblique angle deposition that can produce arrays of nanorods and nanosprings. \cite{6,7,8,9,10,11} In addition, depending on the detailed deposition process, particles can either stick to or bounce off from their impact points, which is determined by a sticking probability, also named ``sticking coefficient'' ($s$). Non-sticking particles are re-emitted and can arrive at other surface points including shadowed valleys. In other words, re-emission has a smoothening effect while shadowing tries to roughen the surface. Both the shadowing and re-emission effects have been proven to be dominant over the surface diffusion and noise, and act as the main drivers of the dynamical surface growth front. \cite{9,10} The prevailing effects of shadowing and re-emission rely on their ``non-local'' character: The growth of a given surface point depends on the heights of near and far-away surface locations due to shadowing and existence of re-emitted particles that can travel over long distances.

\begin{figure}[htb]
\includegraphics[scale=0.5]{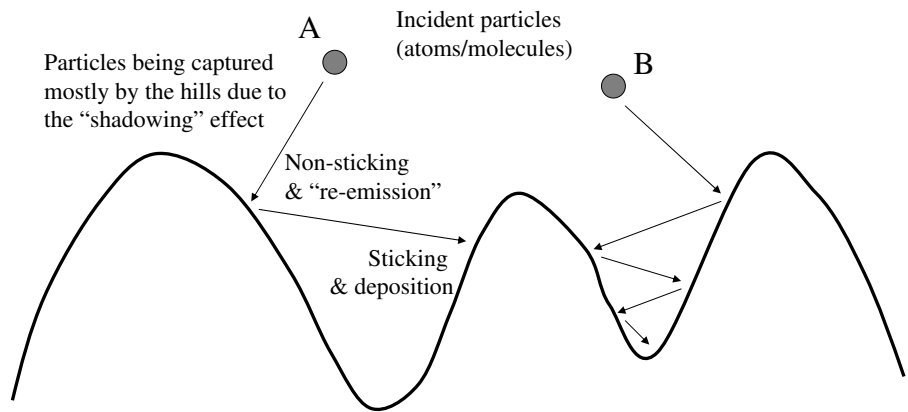}
\caption{Surface of a growing thin film under shadowing and re-emission effects is illustrated.}
\label{fig2}
\end{figure}

Figure \ref{fig3} summarizes some of the experimentally measured sticking coefficient values reported in the literature during evaporation, \cite{24} sputtering, \cite{25,26,27,28,29,30,31,32} and CVD \cite{33,34,35,36,37,38,39} growth of various thin film materials. Names of incident atoms/molecules on the growing film are also labeled. It can be clearly seen from Fig.~\ref{fig3} that incident particles can have sticking probabilities much less than unity in many commonly used deposition systems, which further indicates that re-emission effects should be taken into account in attempts for a realistic thin film growth modeling.

\begin{figure}[htb]
\includegraphics[scale=0.5]{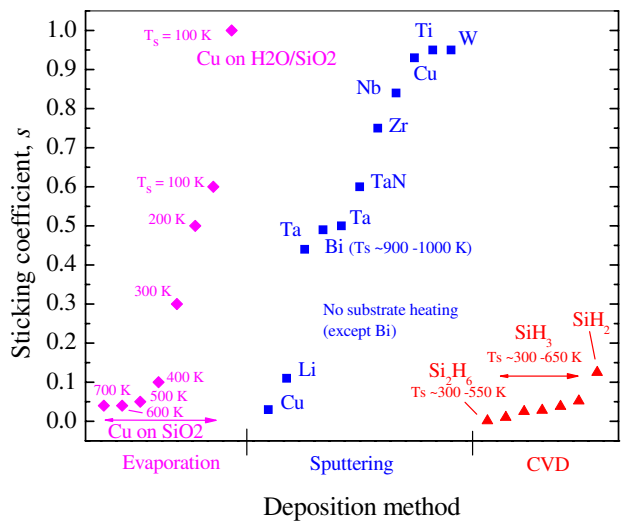}
\caption{(Color online) Some of the experimentally measured sticking coefficient values reported in the literature during evaporation, \cite{24} sputtering, \cite{25,26,27,28,29,30,31,32} and CVD \cite{33,34,3,36,37,38,39} growth are shown. Names of incident atoms/molecules on the growing film are also labeled. In same cases, depositions were done at with substrate heating at temperatures denoted as $T_{s}$ in the figure.}
\label{fig3}
\end{figure}

Due to the complexity of the shadowing and re-emission effects, no growth model has been developed yet within the framework of dynamical scaling theory that take into both these effects and still that can be analytically solved to predict the morphological evolution of thin film or nanostructure deposition. \cite{40} Only recently, shadowing and re-emission effects could be fully incorporated into the Monte Carlo lattice simulation approaches. \cite{10,14,15,16,17,22,23,40,41,42,43} These simulations successfully predicted the experimental results including the $\beta$ values reported in the literature (see Fig.~\ref{fig1}). \cite{20} However, like in experiments, $\beta$ values from simulations ranged all the way from 0 to 1 depending on the sticking coefficients used. For example, Fig.~\ref{fig4} shows $\beta$ values for a Monte Carlo simulated CVD growth obtained for various sticking coefficient and surface diffusion values. \cite{15,17,22,23} It has been observed that re-emission and shadowing effects dominated over the surface diffusion processes due to their long-range non-local character. At small sticking coefficients (e.g. $s<0.5$) re-emission was stronger than the roughening effects of shadowing and Monte Carlo simulations produced smooth surfaces with small $\beta$ values. At higher sticking coefficient values, shadowing effect becomes the dominant process and columnar rough morphologies start to form. \textit{On the other hand, like in experiments, it was not possible to capture a ``universal'' growth behavior using Monte Carlo simulation approaches, which would lead to dynamically common aspects of various thin film growth processes.}

\begin{figure}[htb]
\includegraphics[scale=0.5]{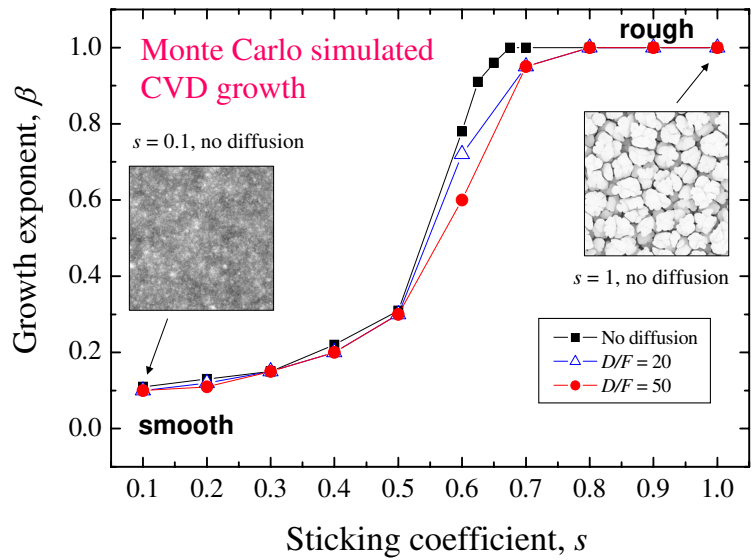}
\caption{(Color online) Growth exponent $\beta$ values for a Monte Carlo simulated chemical vapor deposition (CVD) growth obtained for various first-impact sticking coefficient ($s$) and surface diffusion ($D/F$) values. The sticking coefficient at the second impact after re-emission was set to 1. Two sample surface morphologies are also included for a small $s  = 0.1$ (left) and high $s  = 1$ (right) sticking coefficient value, which leads to a smooth and rough surface topography, respectively.}
\label{fig4}
\end{figure}

Moreover, it has been very recently revealed that \textit{shadowing effect can lead to the breakdown of dynamical scaling theory due the formation of a mounded surface morphology}. \cite{16,17} In these studies, using Monte Carlo simulations it has been shown that for common thin film deposition techniques, such as sputter deposition and CVD, a ``mound'' structure can be formed with a characteristic length scale that describes the separation of the mounds, or ``wavelength'' $\lambda$. It has been found that the temporal evolution of $\lambda$ is distinctly different from that of the mound size, or the lateral correlation length, $\xi$. The formation of the mound structure is due to non-local growth effects, such as shadowing, that lead to the breakdown of the self-affinity of the morphology described by the dynamic scaling theory. The wavelength grows as a function of time in a power law form, $\lambda \sim t^p$, where $p\approx 0.5$ for a wide range of growth conditions while the mound size grows as $\xi\sim t^{1/z}$, where $1/z$ depends on the growth conditions. 

In brief, conventional growth models in dynamic scaling theory can not explain most of the experimental results reported for dynamic thin film growth; and dynamic scaling theory itself often suffers from a breakdown if shadowing effect is present, which is the case for most of the commonly used deposition techniques. On the other hand, simulation techniques were not successful in revealing the possible universal behavior in various growth processes. Furthermore, simulations that can successfully predict the experimental results can not always be easily implemented by a widespread of researchers. Therefore, there is an immense need for alternative and robust modeling approaches for the dynamical growth of thin film surfaces that incorporates easy-to-implement analytical and/or empirical relations which in turn can lead to universal growth behavior aspects of thin films. In this work, we explore a radically new ``network'' modeling approach for the dynamical growth of a large variety of thin film growth systems that can potentially capture universal properties of film growth process and at the same time not suffer from the shortcomings of dynamic scaling theory and Monte Carlo Simulation approaches mentioned above. 

Network modeling pervades various areas of science
ranging from sociology to statistical physics or computer science,
see Ref. 44, and the references therein. A network in
terms of modeling can be defined as a set of items, referred
to as nodes with links connecting them. Examples of real-life
complex networks include the Internet, the World Wide Web,
metabolic networks, transportation networks, social networks,
etc. Recent studies show that many of these networks
share common properties such as having a low degree of
separation among the nodes modeled as Small-World
Networks \cite{45} and having a power-law degree or connectivity
distributions modeled as Scale-Free Networks. \cite{46}

\end{section}

\begin{section}{Origins of Network Behavior during Thin Film Growth}

Interestingly, non-local interactions among the surface points of a growing thin film that originate from shadowing and re-emission effects (Fig.~\ref{fig2}) can lead to non-random preferred trajectories of atoms/molecules before they finally stick and get deposited. For example, during re-emission, the path between two surface points where a particle bounces off from the first and head on to the second can define a ``network link'' between the two. If the sticking coefficient is small, then the particle can go through multiple re-emissions that form links among many more other surface points. In addition, due to the shadowing effect, higher surface points act as the locations of first-capture and centers for re-emitting the particles to other places. In this manner, hills on a growing film resembles to the network ``nodes'' of heavy traffic, where the traffic is composed by the amount of particles re-emitting from the nodes. In terms of network traffic, nodes can be classified as: \textit{source}, \textit{sink}, or \textit{router}. So, the initial point/hill where an atom re-emits can correspond to a ``source'' in a network, and the final point where the atom sticks/settles can be thought as a ``sink'' in the network. Similarly, the intermediate reemission points/hills can be thought as the ``routers''. Therefore, a ``traffic model'' for the thin film growth can then be constructed by counting the number of atoms starting from a point on the film and ending at another point on the film. 

\end{section}

\begin{section}{Monte Carlo Simulations}

Development of network models by our approach requires the track record of the trajectories of re-emitted and deposited atoms/particles. Since it is not possible to experimentally track the trajectories of re-emitted and deposited atoms during dynamic thin film growth, we used 3D Monte Carlo simulation approaches instead, which were already shown to efficiently mimic the experimental processes and predict the correct dynamical growth morphology. \cite{9,10,16,17,22,23,41,42} In these simulations, each incident particle (atom/molecule) is represented with the dimension of one lattice point. As substrate, we used a $N\times N = 512\times512$ size lattice with continuous boundary conditions. A specific angular distribution for the incident flux of particles is chosen depending on the deposition technique being simulated. During normal angle deposition, all the particles are sent from the top along the substrate normal (polar angle $\theta = 0^{\circ}$), while during oblique angle deposition simulations we used a grazing incidence flux where all particles are emitted at a $\theta = 85^{\circ}$ angle from the substrate normal. For CVD, the incident flux had an angular spread according to the distribution function $dP(\theta,\phi)/d\Omega =\cos\theta /\pi$, where $\phi$ is the azimuthal angle. \cite{40} 

At each simulation step, a particle is sent toward a randomly chosen lattice point on the substrate surface.  Depending on the value of sticking coefficient ($s$), the particle can bounce off and re-emit to other surface points. Re-emission direction is chosen according to a cosine distribution centered around the local surface normal. \cite{40} At each impact, sticking coefficient can have different values represented as $s_n$, where $n$ is the order of re-emission ($n = 0$ being for the first impact). In this study, we use a constant sticking coefficient value for all impacts (i.e., $s_{n} = s$ for all $n$) during a given simulation, which is a process also called ``all-order re-emission''. \cite{40} In all the emission and re-emission processes shadowing effect is included, where the particleÕs trajectory can be cut-off by long surface features on its way to other surface points. After the incident particle is deposited onto the surface, it becomes a so called ``adatom''. Adatoms can hop on the surface according to some rules of energy, which is a process mimicking the surface diffusion. However, as noted before, non-local processes of re-emission and shadowing are generally dominant over local surface diffusion effects. Therefore, in this work we did not include surface diffusion in order to better distinguish the effects of re-emission and shadowing effects. After this deposition process, another particle is sent, and the re-emission and deposition are repeated in a similar way. 

In our simulations, deposition time $t$ is represented by number of particles sent to the surface. Final simulation time (total number of particles sent) for all the simulations was $t_{final} = 25\times10^7$. Because of re-emission, deposition rate and therefore average film thickness ($d$) depended on the sticking coefficient $s$ used, and changed with simulation time $t$ approximately according to $d \approx ts /N^{2}$, where lattice size $N$ was 512. 

Furthermore, in our simulations, trajectories of particles during each re-emission process can be tracked in order to reveal the dynamic network behavior in detail. When the simulation time reaches a pre-set value that we called the ``snapshot state'', then we label each particle sent to the surface and start recording the coordinates of lattice point where the particle impacts and also the lattice point where it is re-emitted and makes another impact. Therefore, especially a small sticking coefficient particle can potentially make multiple re-emissions among the surface points and have multiple trajectory data. In order to increase the number of trajectory data for a better statistical analysis while keeping the surface morphology unchanged, we cancelled the final deposition of the particle sent during the trajectory data collection process. In other words, when the simulation time reached the pre-set value, particles were still being sent for re-emission and collection of trajectory data, however, they were not depositing to the surface, therefore not changing the surface morphology. We collected the trajectory data of about 106 re-emitted particles for each snapshot state. We did not include the trajectory data of particles as they re-emitted into the space or if they cross the lattice boundaries, since cross-boundary particles can lead to an artificially long trajectories due to the continuous boundary conditions used. All the simulations results are average of 10 runs (realizations), each time using a different seed number for the random number generator.

\end{section}

\begin{section}{Results and Discussions}

Figure \ref{fig5} shows the snapshot top view images of two surfaces simulated for a CVD type of deposition, at two different sticking coefficients. Figure \ref{fig5} also displays their corresponding particle trajectories projected on the lateral plane. Qualitative network behavior can easily be realized in these simulated morphologies as the trajectories of re-emitted atoms ``link'' various surface points. It can also be seen that larger sticking coefficients [Fig.~\ref{fig5}(b) and (d)] leads to fewer but longer range re-emissions, which are mainly among the peaks of columnar structures. Therefore, these higher surface points act as the ``nodes'' of the system. This is due to the shadowing effect where initial particles preferentially head on hills. They also have less chance to arrive down to valleys because of the high sticking probabilities (see for example particle A illustrated in Fig.~\ref{fig2}). On the other hand, at lower sticking coefficients [Fig.~\ref{fig5}(a) and (c)], particles now go through multiple re-emissions and can link many more surface points including the valleys that normally shadowed by higher surface points (particle B in Fig.~\ref{fig2}). This behavior is better realized in ``surface-degree'' and their corresponding height matrix plots of Fig.~\ref{fig6} measured for CVD grown films at two different sticking coefficients $s = 0.1$ and $s = 0.9$. The high values (darker colors) in surface-degree plots correspond to the highly connected surface sites where these sites get or re-distribute most of the re-emitted particles. At smaller sticking coefficients [Fig.~\ref{fig6}(a)], which leads to a smoother morphology, surface-degree values are quite uniform indicating a uniform re-emission process among hill-to-hills and hill-to-valleys. On the other hand, at high sticking coefficients [Fig.~\ref{fig6}(b)], the high degree nodes are mainly located around the column borders suggesting a dominant column-to-column re-emission. This is consistent with the shadowing effect where columns capture most of the incident particles because of their larger heights, and also their borders are more likely to re-distribute the particles towards the neighboring column sides because of the re-emission process used (i.e., cosine distribution centered along the local surface normal).

\begin{figure}[htb]
\includegraphics[scale=0.5]{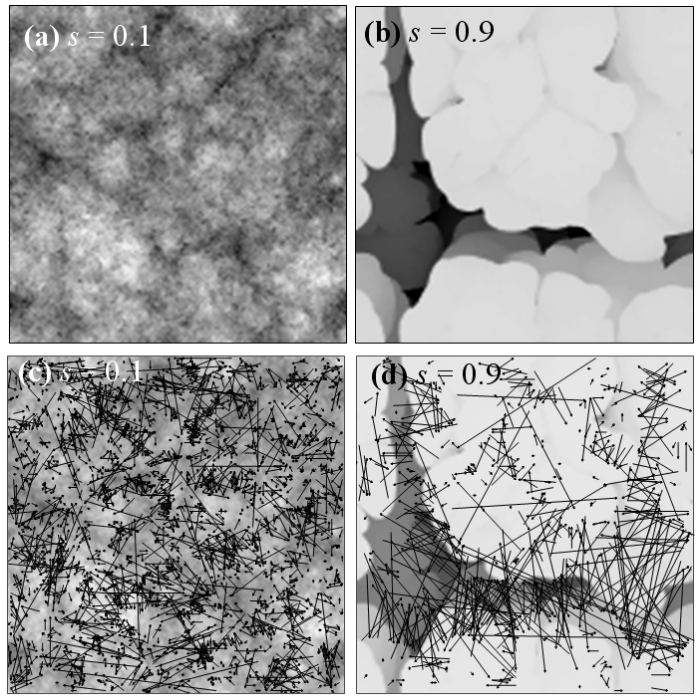}
\caption{Top view images from Monte Carlo simulated thin film surfaces grown under shadowing, re-emission, and noise effects (no surface diffusion is included in these simulations) for sticking coefficients (a) $s = 0.1$ and (b) $s = 0.9$ and with unity sticking coefficient at the second impacts. Each image corresponds to a $128\times128$ portion of the total lattice. The incident flux of particles has an angular distribution designed for chemical vapor deposition (CVD). Corresponding projected trajectories of the re-emitted particles are also mapped on the top view morphologies for (c) $s = 0.1$ and (d) $s = 0.9$. Qualitative network behavior can be seen among surface points linked by the re-emission trajectories.}
\label{fig5}
\end{figure}

\begin{figure}[htb]
\includegraphics[scale=0.5]{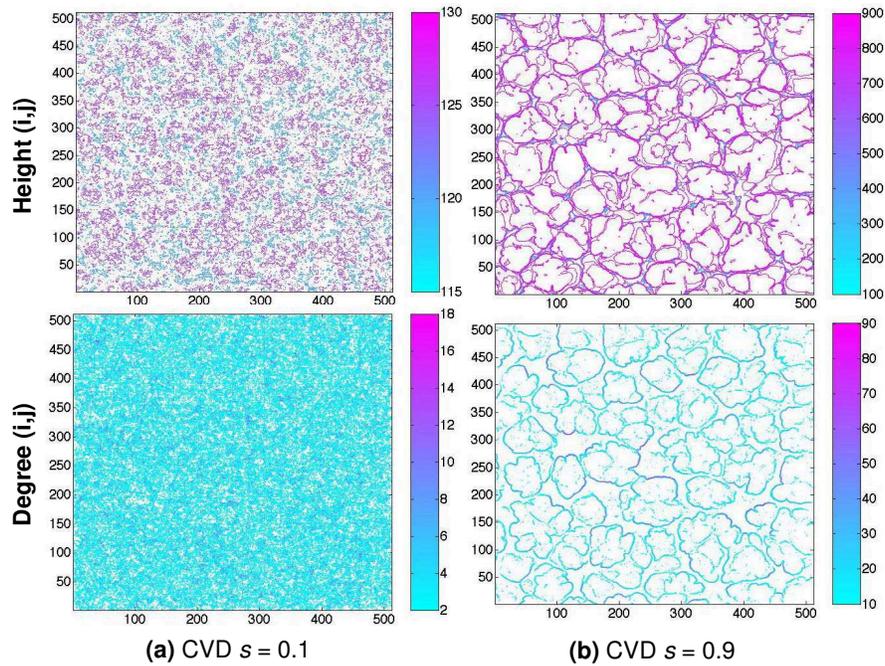}
\caption{(Color online) Height matrix and corresponding surface-degree values are plotted for CVD grown films with sticking coefficients (a) $s = 0.1$ and (b) $s = 0.9$. Total lattice size is $512\times512$ and simulation time for these snapshot states was $t = 23.75\times10^7$ particles.}
\label{fig6}
\end{figure}

Another interesting observation revealed in our Monte Carlo simulations was the dynamic change of network behavior on the trajectories of re-emitted particles. Figure \ref{fig7} shows top view images and their corresponding particle trajectories obtained from the CVD simulations for a sticking coefficient of $s = 0.9$, but this time at different film thicknesses that is proportional to the growth time. The dynamic change in the network topography can be clearly seen: at initial times, when the hills are smaller and more closely spaced, the re-emitted particles travel from one hill to another one or to a valley. However, as the film gets thicker, and some hills become higher than the others and get more separated, particles travel longer ranges typically among these growing hills. The shorter hills that get shadowed become the valleys of the system.

\begin{figure}[htb]
\includegraphics[scale=0.5]{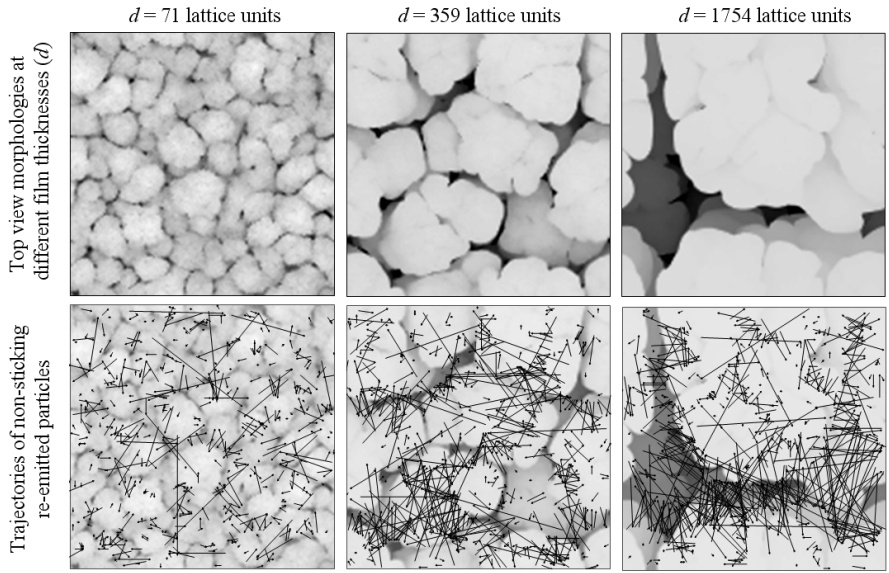}
\caption{First row: Top view images from Monte Carlo simulated thin film surfaces for a CVD growth with $s = 0.9$ at different film thicknesses $d$, which is proportional to growth time. Bottom row: Corresponding projected trajectories of the re-emitted particles qualitatively show the dynamic change in the network topography. }
\label{fig7}
\end{figure}

It is expected that this dynamic behavior should be strongly dependent on the values of sticking coefficients and angular distribution of the incident flux of particles, which determine the strength of re-emission and shadowing effects, respectively. In other words, each deposition technique and material system can have different dynamic network behavior that can lead to various kinds of network systems. For example, as we will show later, the dynamic network among the surface points of a mounded CVD grown film can be quite different than the one among the nanorod and nanospring structures formed in an oblique angle deposition system, where the shadowing effect is most dominant, and also the one during normal angle evaporation, where shadowing effect is almost absent (re-emitted particles during normal angle deposition can still lead to a minimal short-range shadowing effect).

In order to make a more quantitative analysis on the network characteristics of thin film growth dynamics, in Fig.~\ref{fig8}, we plotted the degree distributions $P(k)$ (i.e., proportional to the percentage of surface points having ``degree ($k$)'' number of links through incoming or outgoing re-emitted particles), average distance $\left<l_{k}\right>$ versus degree k (i.e., the average ``lateral'' distance particles travel that are re-emitted from/to surface sites having $k$ number of links), and distance distributions $P(l)$ (i.e., proportional to the probability of a re-emitted particle traveling lateral distance $l$) for Monte Carlo simulated normal incidence evaporation, oblique angle deposition, and CVD thin film growth for various sticking coefficients. The left and right columns in Fig.~\ref{fig8} correspond to the initial (thinner films) and later (thick films) stages of the growth times, respectively. First, the comparison of degree distributions [Fig.~\ref{fig8}(a) and (b)] of normal incidence and oblique angle growth reveals that independent of the most sticking coefficients used and also their growth time, universal behavior exists for both deposition techniques: There is an exponential degree distribution for normal angle evaporation (confirmed in the semi-log plots, not shown here), while this behavior is mainly power-law for oblique angle deposition with an exponential tail. Interestingly, quantitative values of degree distributions for both normal and oblique angle depositions also seem to be independent of the sticking coefficient used, which becomes clearer at later stages of the growth [Fig.~\ref{fig8}(b)], leading to two distinct distributions for each deposition. The power-law observed in degree distribution of oblique angle deposition has a $P(k) \sim k^{-2}$ behavior apparent at later stages. All these suggest the possibility of a universal behavior in normal and oblique angle growth independent of the sticking coefficient. This is quite striking since each different sticking coefficient corresponds to a different type of morphological growth (i.e., smoother surfaces for smaller sticking coefficients and rougher surfaces for higher sticking coefficients), yet the degree distribution in network traffic of re-emitted particles seems to reach a unique universal state.

\begin{figure}[htb]
\includegraphics[scale=1]{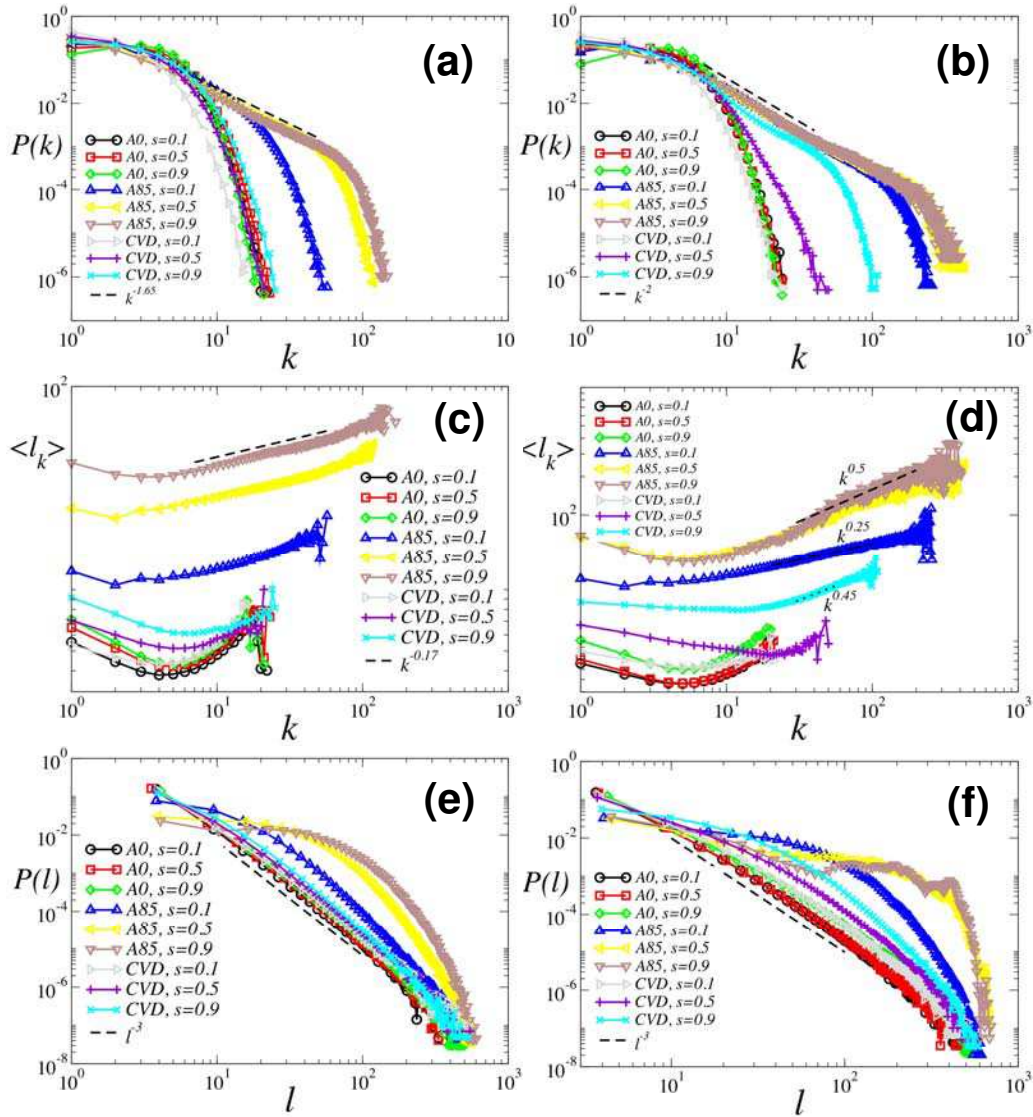}
\caption{(Color online) Behavior of degree distributions $P(k)$ (top row), average distance $\left<l_k\right>$ versus degree (middle row), and distance distributions $P(l)$ (bottom row) for network models of a Monte Carlo simulated normal incidence evaporation (A0), oblique angle deposition (A85), and CVD thin film growth for various sticking coefficients $s$ and for two different deposition time $t$ (left column: $t = 1.25\times10^{}7$ particles, and right column: $t = 23.75\times10^7$ particles) are shown.}
\label{fig8}
\end{figure}

As can be seen in Fig.~\ref{fig8}(a) and (b), the re-emission process which is the dominant process in normal angle growth promotes an exponential degree distribution; while shadowing which is the governing effect during oblique angle deposition leads to a power-law distribution. On the other hand, CVD shows an exponential degree distribution at initial times of the growth, while it becomes closer to power-law type for higher sticking coefficients $s > 0.5$. This is believed to be competing forces of re-emission and shadowing effects, where the re-emission is more dominant for smaller sticking coefficients and at initial times of the growth when the film is smoother, leading to an exponential degree distribution. However, shadowing effect originating from the obliquely incident particles within the angular distribution of CVD flux can lead to a power-law behavior at higher sticking coefficients especially when the film gets rougher at later stages of the growth. A power-law degree distribution corresponds to a more correlated network that is consistent with the long-range, column-to-column traffic observed in surface-degree plots of high sticking coefficient CVD above [Fig.~\ref{fig6}(b)]. It is also realized that especially for high sticking coefficients, there exist high degree nodes represented with data points at the tails of the degree distributions. These relatively small percentage but highly connected nodes are mainly located at the column edges as seen in surface degree plot of Fig.~\ref{fig6}(b) and are likely to be the ``hubs'' of the network. Therefore, briefly, degree distribution during CVD growth can be similar to the universal line of normal incidence growth for smaller sticking coefficients ($s < 0.5$) showing an exponential behavior with a short range network traffic; or it can converge to the universal power-law degree distribution of oblique angle deposition for higher sticking coefficients ($s > 0.5$) leading to a highly correlated network driven mostly at column edges.

In addition, it is revealed from average distance versus degree plots of Fig.~\ref{fig8}(c) and (d) that nodes with high degree are mainly linked with long-distance surface points. Independent of the deposition method, the average distance changes with degree k according to a power law behavior, where the value of exponent increases as the flux become more oblique [i.e., A0 $\to$ A85 $\to$ CVD in Fig.~\ref{fig8}(c) and (d)], sticking coefficient increases, and the film gets thicker [i.e., Fig.~\ref{fig8}(c) and (d)]. In other words, when shadowing effect becomes more dominant and film morphology gets more columnar, high degree nodes can exchange atoms with longer distance surface points. This also implies that high degree nodes placed on column edges [Fig.~\ref{fig6}(b)] of high sticking coefficient growth are more likely to transfer particles with far away other column edges as well. This process is further supported by the distance distribution plots of Fig.~\ref{fig8}(e) and (f) where as the sticking coefficient is increased (i.e., less re-emission) and more obliquely deposited particles are introduced (more shadowing effect), more percentage particles start to travel longer distances. On the other hand, for smaller sticking coefficients and normal angle deposition, where the re-emission effect is more dominant, average distances particles travel from high degree nodes become significantly less compared to high sticking coefficient and oblique angle depositions. This suggests networking during re-emission dominated growth occurs mainly among smaller size hills and valleys, consistent with the surface-degree plot of Fig.~\ref{fig6}(a). 

A more interesting universal behavior is observed in a ``weighted and scaled average distance'' versus degree plots of Fig.~\ref{fig9}(a) and (b). Here we re-scale average distance for nodes with degree $k$, $\left<l_{k}\right>$, first with degree $k$, then with the average distance value of all nodes $\left<l\right>$ (average distance of all links), and plot $\left<l_{k}\right>/(k\left<l\right>)$ versus $k$. After re-scaling, independent of the deposition technique used, sticking coefficients, and the growth time, all curves fall on a similar line obeying a power-law behavior with $\left<l_{k}\right>/(k\left<l\right>) \sim k^{-1.2}$. The origin of the -1.2 value of the exponent is not clear and is under investigation. 

\begin{figure}[htb]
\includegraphics[scale=0.65]{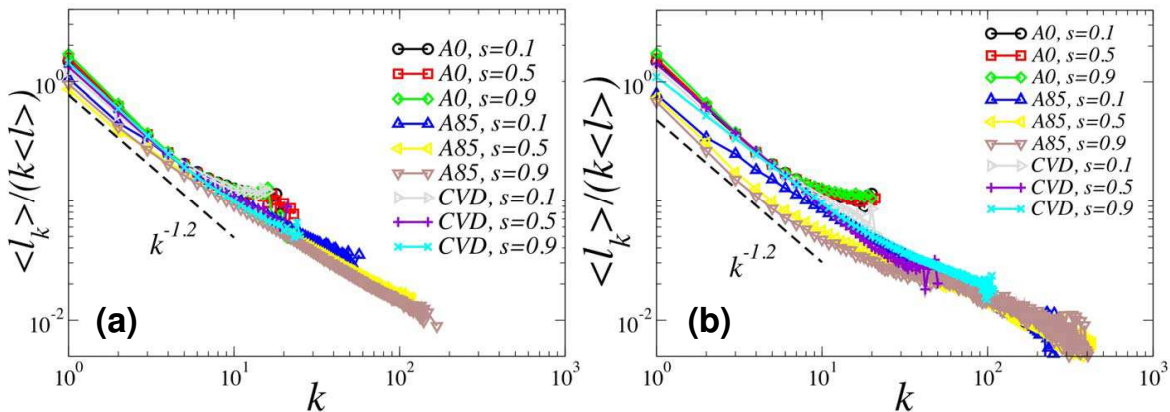}
\caption{(Color online) Weighted average distance $\left<l_k\right>/(k\left<l\right>)$ versus degree $k$ for network models of a Monte Carlo simulated normal incidence evaporation (A0), oblique angle deposition (A85), and CVD thin film growth for various sticking coefficients $s$ and for two different deposition time $t$ [(a) $t = 1.25\times10^7$ particles, and (b) $t = 23.75\times10^7$ particles] are shown.}
\label{fig9}
\end{figure}

Another universal behavior is observed in distance distribution plots: Independent of sticking coefficients, normal incidence growth shows a power-law behavior with $P(l) \sim l^{-3}$. A similar power-law behavior with an exponent of -2.75 has been observed in the distance distribution plots during a normal incidence growth simulation with re-emission (p.83 of Ref.14). The authors of that work did not use a snapshot state approach, surface morphology continuously changed, and therefore they measured a kind of average distance distribution of the whole growth simulation. However, their exponent value is still close to our results and agrees with our findings that dynamic network behavior during normal incidence deposition does not change significantly due to the relatively smooth morphology throughout the growth.  On the other hand the behavior in distance distribution plots is exponential for oblique angle deposition (confirmed in the semi-log plots, not shown here). CVD has a power law behavior similar to that of normal incidence growth with $P(l) \sim l^{-3}$ at smaller coefficients and at initial times of the growth, and becomes exponential similar to oblique angle deposition at higher sticking coefficients apparent especially later stages of the growth.

\end{section}


\begin{section}{Conclusion}

In conclusion, we presented a new network modeling approach for various thin film growth techniques that incorporates re-emitted particles due to the non-unity sticking coefficients. We define a network link when a particle is re-emitted from one surface site to another one. Monte Carlo simulations are used to grow films and dynamically track the trajectories of re-emitted particles. We performed simulations for normal incidence, oblique angle, and CVD techniques. Each deposition method leads to a different dynamic evolution of surface morphology due to different sticking coefficients involved and different strength of shadowing effect originating from the obliquely incident particles. Traditional dynamic scaling analysis on surface morphology cannot point to any universal behavior. On the other hand, our detailed network analysis reveals that there exist universal behaviors in degree distributions, weighted average degree versus degree, and distance distributions independent of the sticking coefficient used and sometimes even independent of the growth technique. We also observe that network traffic during high sticking coefficient CVD and oblique angle deposition occurs mainly among edges of the columnar structures formed, while it is more uniform and short-range among hills and valleys of small sticking coefficient CVD and normal angle depositions that produce smoother surfaces.

\end{section}



\begin{thebibliography}{99}

\bibitem{1}
D. L. Smith, \textit{Thin-Film Deposition: Principles and Practice} (McGraw-Hill, New York, 1995).
\bibitem{2}
P. Meakin, \textit{Fractals, Scaling, and Growth Far from Equilibrium} (Cambridge University Press, Cambridge, England, 1998).
\bibitem{3}	
A.-L. Barabasi and H. E. Stanley, \textit{Fractal Concepts in Surface Growth} (Cambridge University, Cambridge, England, 1995).
\bibitem{4}	
N. O. Young and J. Kowal, Nature (London) \textbf{183}, 104 (1959).
\bibitem{5}	
T. Motohiro and Y. Taga, Appl. Opt. \textbf{28}, 2466 (1989).
\bibitem{6}	
R. M. Azzam, Appl. Phys. Lett. \textbf{61}, 3118 (1992).
\bibitem{7}	
K. Robbie, M. J. Brett, and A. Lakhtakia, Nature (London) \textbf{384}, 616 (1996).
\bibitem{8}	
K. Robbie, G. Beydaghyan, T. Brown, C. Dean, J. Adams, and C. Buzea, Rev. Sci. Instrum. \textbf{75}, 1089 (2004).
\bibitem{9}
T. Karabacak and T.-M. Lu, in \textit{Handbook of Theoretical and Computational Nanotechnology}, edited by M. Rieth and W. Schommers (American Scientific, Stevenson Ranch, CA, 2005), Chap. 69, p. 729. 
\bibitem{10}
T. Karabacak, G.-C. Wang, and T.-M. Lu, J. Vac. Sci. Technol. A \textbf{22}, 1778 (2004).
\bibitem{11}
A. Lakhtakia and R. Messier, \textit{Sculptured Thin Films: Nanoengineered Morphology and Optics} (SPIE Press, Bellingham, WA, 2005).
\bibitem{12}
F. Family and T. Vicsek, J. Phys. A \textbf{18}, L75  (1985).
\bibitem{13}
F. Family, J. Phys. A \textbf{19}, L441 (1986).
\bibitem{14}
M. Pelliccione and T.-M Lu, \textit{Evolution of Thin Film Morphology: Modeling and Simulations} (Springer, New York, 2007).
\bibitem{15}
T.-M. Lu, Y.-P. Zhao, J.T. Drotar, T. Karabacak, and G.-C. Wang, \textit{Morphological and Compositional Evolution of Thin Films}, MRS Symposia Proceedings No. 749 (Materials Research Society, Pittsburgh, 2003) p. 3.
\bibitem{16}
M. Pelliccione, T. Karabacak, and T.-M. Lu, Phys. Rev. Lett. \textbf{96}, 146105 (2006).
\bibitem{17}
M. Pelliccione, T. Karabacak, C. Gaire, G.-C. Wang, and T.-M. Lu, Phys. Rev. B \textbf{74}, 125420 (2006).
\bibitem{18}
Y.-P. Zhao, G.-C. Wang, and T.-M. Lu, \textit{Characterization of Amorphous and Crystalline Rough Surfaces: Principles and Applications} (Academic Press, New York, 2001).
\bibitem{19}
F. Family and T. Viscek, \textit{Dynamics of Fractal Surfaces} (World Scientific, Singapore, 1991).
\bibitem{20}
Survey of experimental $\beta$ values is obtained by updating the data presented in Ref.~15.
\bibitem{21}
R.P.U. Karunasiri, R. Bruinsma, and J. Rudnick, Phys. Rev. Lett. \textbf{62}, 788 (1989).
\bibitem{22}
T. Karabacak, Y.-P. Zhao, G.-C. Wang, T.-M. Lu, Phys. Rev. B \textbf{64}, 085323 (2001).
\bibitem{23}
T. Karabacak, Y.-P. Zhao, G.-C. Wang, T.-M. Lu, Phys. Rev. B \textbf{66}, 075329 (2002).
\bibitem{24}
X. Xu and D. W. Goodman, Appl. Phys. Lett. \textbf{61}, 1799 (1992).
\bibitem{25}
E. M. van Veldhuizen and F. J. de Hoog, J. Phys. D \textbf{17}, 953 (1984).
\bibitem{26}
A. Bogaerts, J. Naylor, M. Hatcher, W. J. Jones, and R. Mason, J. Vac. Sci. Technol. A \textbf{16}, 2400 (1998).
\bibitem{27}
K. Obara, Z. Fu, M. Arima, T. Yamada, T. Fujikawa, N. Imamura, and N. Terada, J. of Crystal Growth \textbf{237--239}, 2041 (2002).
\bibitem{28}
S. Migita, K. Sakai, H. Ota, Z. Mori, and R. Aoki, Thin Solid Films \textbf{281-282}, 510 (1996).
\bibitem{29}
A. Bogaerts, E. Wagner, B. W. Smith, J. D. Winefordner, D. Pollmann, W. W. Harrison, and R. Gijbels, Spectrochim. Acta Part B \textbf{52}, 205 (1997). 
\bibitem{30}
A. J. Toprac, B. P. Jones, J. Schlueter, and T. S. Cale, \textit{Evolution of Thin Films and Surface Structure and Morphology}, MRS Symposia Proceedings No. 355 (Materials Research Society, Pittsburgh, 1995), p. 575.
\bibitem{31}
O. Yamazaki, K. Iyanagi, S. Takagi, and K. Nanbu, Jpn. J. Appl. Phys., Part 1  \textbf{41}, 1230 (2002).
\bibitem{32}
D. Liu, S. K. Dew, M. J. Brett, T. Smy, and W. Tsai, J. Appl. Phys. \textbf{75}, 8114 (1994).
\bibitem{33}
R. J. Buss, P. Ho, W. G. Breiland, and M. E. Coltrin, in \textit{Deposition and Growth: Limits for Microelectronics}, AIP Conf. Proc. No. 167, edited by G. W. Rubloff, (AIP, New York, 1988), p. 34.
\bibitem{34}
R. J. Buss, P. Ho, W. G. Breiland, and M. E. Coltrin, J. Appl. Phys. \textbf{63}, 2808 (1988).
\bibitem{35}
C. C. Tsai, J. G. Shaw, B. Wacker, and J. C. Knights, \textit{Amorphous Silicon Semiconductors--Pure and Hydrogenated Symposium},  MRS Symposia Proceedings No. 95 (Materials Research Society, Pittsburgh, 1987), p. 219.
\bibitem{36}
J. Perrin and T. Broekhuizen,  \textit{Photon, Beam, and Plasma Stimulated Chemical Processes at Surfaces}, MRS Symposia Proceedings No. 75 (Materials Research Society, Pittsburgh, 1987), p. 210.
\bibitem{37}
J. Perrin and T. Broekhuizen, Appl. Phys. Lett. \textbf{50}, 433 (1987).
\bibitem{38}
R. Robertson and A. Gallagher, J. Appl. Phys. \textbf{59}, 3402 (1986).
\bibitem{39}
J. Robertson, J. Non-Cryst. Solids \textbf{266--269}, 79 (2000).
\bibitem{40}
J. T. Drotar, Y.-P. Zhao, T.-M. Lu, and G.-C. Wang, Phys. Rev. B \textbf{61}, 3012 (2000).
\bibitem{41}
T. Karabacak, J.P. Singh, Y.-P. Zhao, G.-C. Wang, and T.-M. Lu, Phys. Rev. B \textbf{68}, 125408 (2003).
\bibitem{42}
T. Karabacak, G.-C. Wang, and T.-M. Lu, J. Appl. Phys. \textbf{94}, 7723 (2003).
\bibitem{43}
T. Smy, D. Vick, M. J. Brett, S. K. Dew, A. T. Wu, J. C. Sit, and K. D. Harris, J. Vac. Sci. Technol. A \textbf{18}, 2507 (2005).
\bibitem{44}
S. Boccaletti, V. Latora, Y. Moreno, M. Chavez, and D.-U. Hwang, Phys. Rep. \textbf{424}, 175 (2006).
\bibitem{45}
D. J. Watts and S. H. Strogatz, Nature (London) \textbf{393}, 440 (1998).
\bibitem{46}
A.-L. Barabasi and R. Albert, Science \textbf{286}, 509 (1999).

\end{thebibliography}
\end{document}